\documentclass[]{article}

\usepackage{xcolor}
\usepackage{amsmath}

\usepackage[affil-it]{authblk}

\usepackage{graphicx}

\begin{document}

%
%

\title{Towards the azimuthal characteristics of ionospheric and seismic effects of "Chelyabinsk" meteorite fall according to the data from coherent radar, GPS and seismic networks}

%
%

%
%

\author{O.I. Berngardt,
N.P. Perevalova,
K.A. Kutelev,
G.A. Zherebtsov}
\affil{Institute of Solar-Terrestrial Physics SB RAS, 126a, Lermontov st., Irkutsk, 664033, Russia}

\author{A.A. Dobrynina}
\affil{Institute of the Earth Crust SB RAS,  128, Lermontov st. Irkutsk, 664033, Russia}

\author{N.V. Shestakov}
\affil{Far Eastern Federal University, 8, Sukhanov st., Vladivostok, 690950, Russia}
\affil{Institute of Applied Mathematics, FEB RAS, 7, Radio st., Vladivostok, 690041, Russia}

\author{R.V.Zagretdinov}
\affil{Federal Kazan University, 18, Kremlyovskaya st., Kazan, 420008, Russia}
\author{V.F.Bakhtiyarov}
\affil{RPC "Geopoligon" Federal Kazan University, Kazan, Russia}
\author{O.A.Kusonsky}
\affil{Institute of Geophysics, UrB RAS, 100, Amundsen st.   Ekaterinburg, 620016, Russia}


\maketitle









%
%


\begin{abstract}

We present the results of a study of the azimuthal characteristics of ionospheric 
and seismic effects of the meteorite 'Chelyabinsk', based on the data from the network 
of  GPS receivers, coherent decameter radar EKB SuperDARN and network of seismic 
stations, located near the meteorite fall trajectory. It is shown, that 6-14 minutes 
after the bolide explosion, GPS network observed the cone-shaped wavefront of TIDs 
that is interpreted as a ballistic acoustic wave. The typical TIDs propagation 
velocity were observed $661 \pm 256 m/s $, which corresponds to the expected 
acoustic wave speed for 240km height.

14 minutes after the bolide explosion, at distances of 200km we observed the 
emergence and propagation of a TID with spherical wavefront, that is interpreted 
as  gravitational mode of internal acoustic waves. The propagation velocity of 
this TID was $ 337 \pm 89 m/s $ which corresponds to the propagation velocity 
of these waves in similar situations. At EKB SuperDARN radar, we observed TIDs 
in the sector of azimuthal angles close to the perpendicular to the meteorite 
trajectory. The observed TID velocity (400 m/s) and azimuthal properties 
correlate well with the model of ballistic wave propagating at 120-140km altitude.

It is shown, that the azimuthal distribution of the amplitude of vertical seismic 
oscillations can be described qualitatively by the model of vertical strike-slip 
rupture, propagating at 1km/s along the meteorite fall trajectory to distance of 
about 40km. These parameters correspond to the direction and velocity of propagation 
of the ballistic wave peak by the ground.

It is shown, that the model of ballistic wave caused by supersonic motion and burning 
of the meteorite in the upper atmosphere can satisfactorily explain the various 
azimuthal ionospheric effects, observed by the coherent decameter radar EKB SuperDARN, 
GPS-receivers network, as well as the azimuthal characteristics of seismic waves at 
large distances.

\end{abstract}

%
%

%


%
%

\section{Introduction}

The bolide, observed on February, 15 2013 above Chelyabinsk region, Russia, 
is the second powerful event since the fall of the Tunguska meteorite in 1908 
\cite{Zotkin_1966, BenMenahem_1975}. According to \cite{Popova_2013},   
'Chelyabinsk' meteorite had an initial size of about 19.8 meters and an 
initial weight of about 13000 tons. When entering the atmosphere, its 
velocity was about 18.6 km/s. The most powerful explosion (the main 
explosive disruption) occurred at 03:20:32UT (09:20:32LT) at $\approx 27 km$  
height. After that, two more explosions occurred: second disruption (03:20:33.4UT) 
and third disruption (03:20:34.7UT) \cite{Popova_2013}.The energy of its burning 
in the atmosphere, that caused the destruction of buildings and windows, according 
to the calculations of \cite{Popova_2013} was equivalent to 590 kilotons (kT) of TNT, 
while the part of the energy, associated with explosions, looks relatively small 
\cite{Popova_2013} (see support materials(SM), pp.62-69 in \cite{Popova_2013}). The paper \cite{Alpatov_2013} 
has shown, that the kinetic energy of the explosion of the 'Chelyabinsk' meteorite was 
$\approx 20kT$ based on the data of  11 infrasound stations of International Monitoring 
System of nuclear tests (MSM). 

Midscale travelling ionospheric disturbances (MSTIDs) with few hundred 
kilometers spatial scale, and  few tens of minutes temporal scale, that 
were caused by the flight and explosion of the 'Chelyabinsk' meteorite, 
were detected by various instruments \cite{Alpatov_2013, Givishvili_2013, 
Gokhberg_2013, Berngardt_2013a, Yang_2014, Ruzhin_2014,Chernogor_2015}.

By using low-orbital tomography at vertical ionospheric profiles over the 
European territory of Russia, wave-like disturbances (wavelength $ \approx 1.2^o $ at latitude) 
were detected, that appear 2.5-3 hours after the explosion \cite{Alpatov_2013, Givishvili_2013}. 
The disturbances of the total electron content (TEC), caused by the 
explosion, were detected by ARTU GPS station \cite{Gokhberg_2013, Ruzhin_2014, Chernogor_2015}. 
According to \cite{Gokhberg_2013} TEC variations had an inverted N-wave shape. 
The distribution of TEC variations amplitudes was not spherically symmetric. 
The ionospheric response to the explosion had a number of features (waveform, 
axial symmetry in the ionospheric disturbances distribution)  distinguishing 
it from the ionospheric response to earthquakes, tsunamis, explosions and 
volcanic eruptions \cite {Gokhberg_2013}. 

In \cite{Yang_2014}, 
using GPS networks near the meteorite trajectory, in Japan and in the United 
States, three different types of MSTIDs were found: the higher-frequency 
(4.0-7.8mHz, periods 4.2-2.1min) disturbances were observed around ARTU 
station and having $\approx 862m/s$ velocity; the lower frequency (1.0-2.5mHz, 
periods 16.7-6.7min) disturbances, propagating with a mean velocity $\approx 362m/s$, 
were observed at distances of 300-1500km from Chelyabinsk; the higher-frequency 
(2.7-11mHz, periods 1.5-6.2min) disturbances with a mean propagation velocity 
$\approx 733m/s$ were registered in USA region. Within 14min after the bolide 
explosion, on several nearby GPS-stations different TEC variations were observed: 
some of them were caused by solar terminator and  some of them propagate radially 
from the explosion epicenter to the distances up to 500-600km with 320-360m/s 
velocity \cite{Berngardt_2013a}. At 06:00-10:00UT at most GPS stations, located 
away from the meteorite trajectory, an intensive TEC variations were observed, which 
had the form of wave packets with 30-40min period \cite{Berngardt_2013a}.

According to the over-the-horizon coherent EKB SuperDARN radar, located 200 km 
from the meteorite explosion place, it is shown that the flight and 
fall of the meteorite caused the formation of MSTIDs with 150-200km horizontal 
spatial scale, propagating throughout the entire ionosphere (in E- and F-layers) 
to the distances up to 1500km from the epicenter. They have two typical propagation 
velocities  - 220m/s and 400m/s \cite{Berngardt_2013b}.

This paper presents the results of studying the azimuthal features of MSTIDs and 
seismic effects of "Chelyabinsk" meteorite according to GPS network, EKB SuperDARN 
radar  and network of seismic stations, located near the meteorite trajectory.

\section{Observations}
\subsection{GPS TEC variations}

\begin{figure}
\includegraphics[scale=0.15]{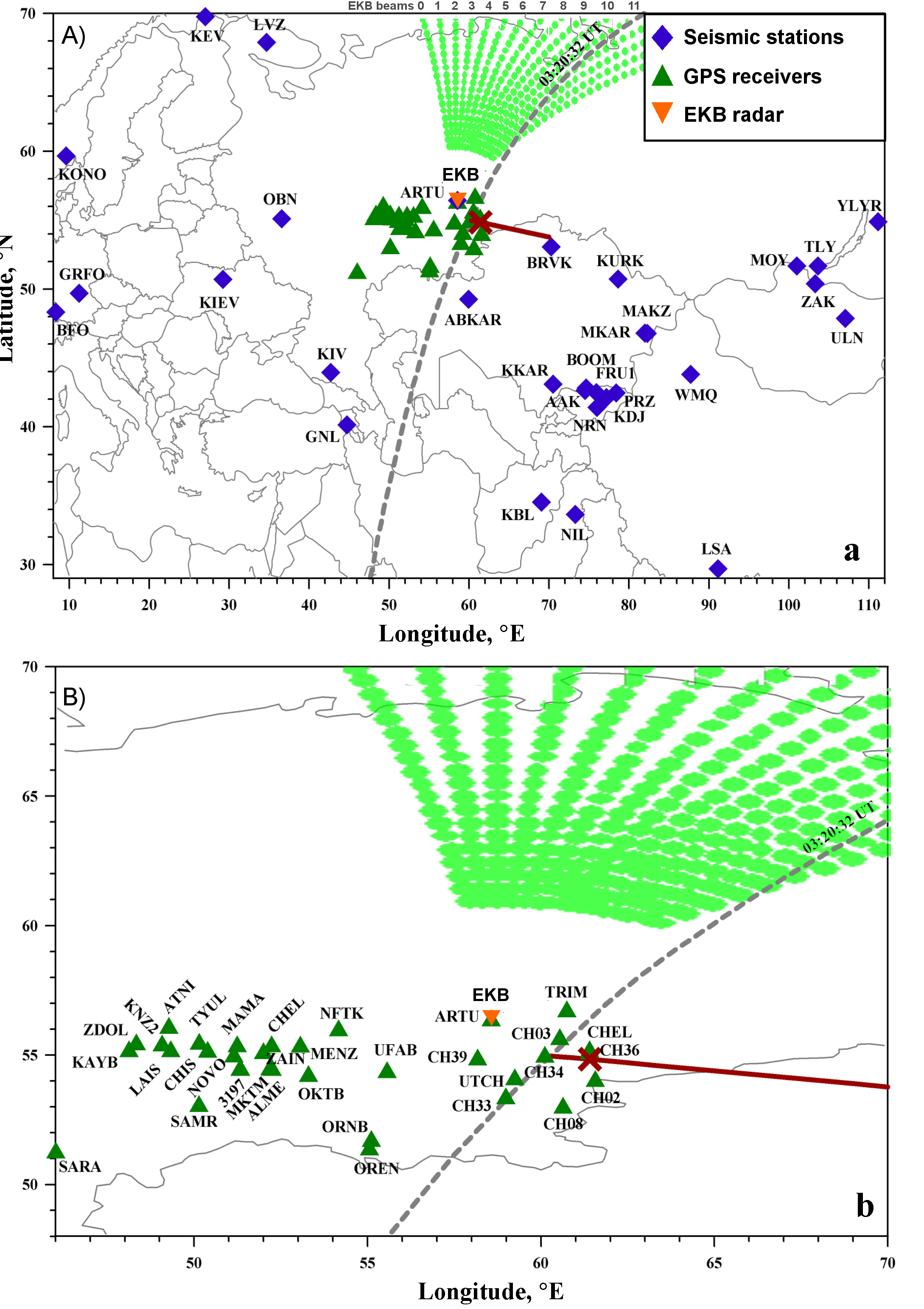}
\caption{
Measurements scheme during the meteorite 'Chelyabinsk' fall. The gray dashed line marks 
the position of the solar terminator for 03:20:32UT. Thick straight line shows the 
meteorite trajectory according to \cite{Popova_2013}, the cross marks the strongest 
explosion.
}
\label{fig:1} 
\end{figure}

The study of TEC variations in the ionosphere during Chelyabinsk meteoroid flight 
was based on phase measurements of double-frequency GPS receivers (Fig.\ref{fig:1}). 
The data (with 1-sec sampling rate) of ARTU GPS-station included in the International 
GNSS Service (IGS) network were downloaded from the Scripps Orbit and Permanent Array 
Center website (http://sopac.ucsd.edu). GPS data (with 30-sec sampling rate) of 
stations TRIM and ORNB operated by NAVGEOKOM company were available from http://www.navgeocom.ru. 
Data of GPS receivers in  the Chelyabinsk region (CHEL, CH02, CH03, CH08, CH33, CH34, CH36, CH39; 
 5-sec sampling rate) were kindly provided by LLC "GEOSalyut" (Moscow) and LLC "Poleos" (Chelyabinsk). 
 The Kazan GPS network data (stations LME, ATNI, CHEL, CHIS, KAYB, KZN2, LAIS, MAMA, MENZ, MKTM, NFTK, 
 NOVO, OKTB, OREN, SAMR, SARA, TYUL, UFAB, UTCH, ZAIN, ZDOL,  30-sec sampling rate; stations 3192, 3196, 
 3197, 3199, 3204, 15-sec sampling rate) were provided by Federal Kazan University (KFU) and 
 RPC "Geopoligon" KFU (Kazan). The last five GPS-stations are very close to each other and 
 shown as single 3197 station at Fig.\ref{fig:1}.

The area around F2-layer maximum (height $h_{max}$) make the main contribution to TEC variations. 
Therefore, it is assumed that TEC variations are formed in an ionospheric piercing point (IPP), where 
IPP is the point of intersection of  the "receiver-satellite" line of sight (LOS) with an ionospheric 
layer at $h_{max}$. According to ARTI ionosonde, located in Ekaterinburg, during meteorite fall $h_{max}$
was about 243 km. We used this value in our calculations.

To separate disturbances triggered by the meteoroid explosion, the initial TEC series I(t) were converted 
to an equivalent "vertical" value \cite{Klobuchar_1986}, and then we filtered it in the period range 1-20 
min \cite{Afraimovich_1998}. TEC variations occurring on the day of the  meteorite fall were compared with 
the TEC behavior observed on previous and subsequent days. The analyzed period featured quiet geomagnetic 
conditions: Kp-index did not exceed 1 within 00:00-12:00 UT, and there were no strong solar flares. The 
seismic background was also quiet in the Chelyabinsk region and nearby areas. Some difficulties in separating 
the meteorite-generated ionospheric disturbances were caused by the fact that the meteoric impact occurred at 
sunrise, when the ionosphere exhibited strong variability. \cite{Afraimovich_2009} demonstrated that solar 
terminator (ST) generates TEC wave disturbances in the ionosphere. Such disturbances can be seen for 2-4 h 
and even precede ST, which is attributed to the ST passage through the magnetoconjugate region. 
Between 02:00 and 06:00 UT on February 14-16, 2013, almost in all LOSs in the TEC variations there were 
quite intense oscillations with an amplitude of ~0.2 TECU and a period of 15-17 min. It is most likely 
that they were caused by the ST passage. 

\begin{figure}
\includegraphics[scale=0.6]{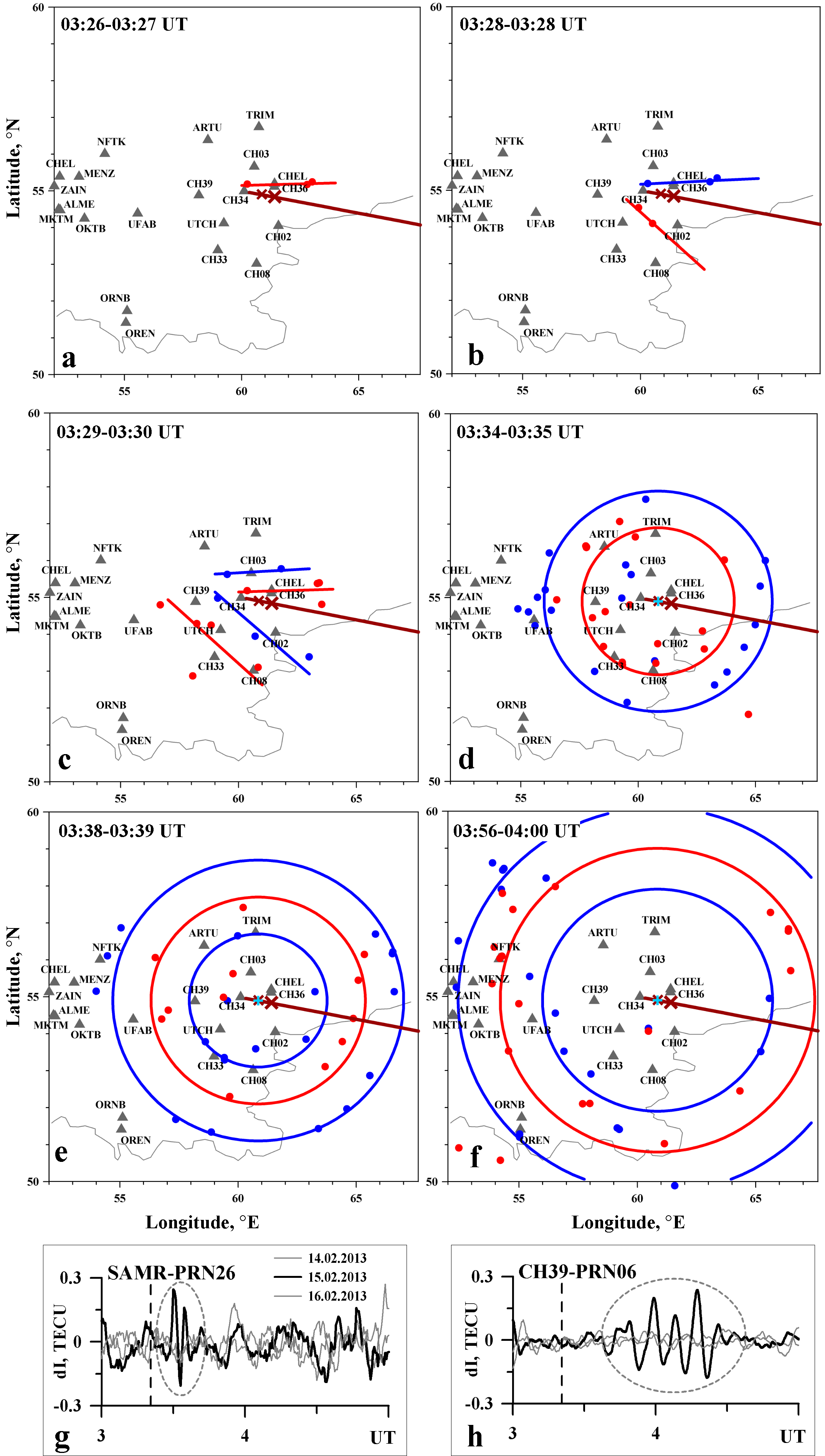}
\caption{
Propagation of the TEC disturbances (a-f). The meteorite trajectory is shown according 
to \cite{Popova_2013} by brown straight line. Three main disruption episodes are 
marked by brown crosses as given in \cite{Popova_2013}. The largest brown cross 
displays the position of main explosive disruption (the airburst). Red dots corresponds 
to TEC maxima positions, blue dots corresponds to TEC minima positions. TEC disturbances 
registered at GPS-station SAMR (satellite PRN26) (g), CH39 (satellite PRN06) (h). The 
airburst time is marked by a vertical dashed line. 
}
\label{fig:2} 
\end{figure}

Nevertheless, in spite of the presence of ST-caused TEC oscillations, in some receiver-satellite LOSs we 
managed to separate disturbances with a characteristic shape corresponding to the shape of a shock acoustic 
wave. In Fig.\ref{fig:2}g-h, such disturbances $\Delta TEC$ are marked with dashed lines. These disturbances 
were oscillations with a period of $\approx 10$ min and amplitude 0.07-0.5 TECU exceeding the level of background 
oscillations on the reference days.

To draw a spatial-temporal picture of the TEC response to the meteoroid explosion, we have constructed TEC 
disturbance maps (Fig.\ref{fig:2}a-f). Figure 2a-f presents spatial distribution of minima (blue dots) and 
maxima (red dots) of TEC oscillations (at $h_{max}=243km $) over a period 03:26-04:00 UT. Thick concentric 
lines indicate the position of the disturbance wave front which can be traced through TEC variations 
minima (blue) and maxima (red). The first TEC disturbances were registered ~6 min after the meteorite 
explosion at distances of 80-100 km from the airburst location. Over a period 03:26-03:33 UT,  the TEC 
disturbances are characterized by fast dynamics. The front of disturbances has a form of the cone 
dispersing from the meteorite trajectory. The angle at a cone peak changes from $4.9^o$ to $22.9^o$ 
(the average angle value is about $14.3^o$). The average velocity was estimated from movement of front 
lines: $V=661\pm 256 m/s$. Since 03:34 UT the behavior of the TEC disturbances corresponds to the spherical 
wave front propagating from the center with coordinates $54.90^oN; 60.85^oE$ (blue cross in Fig.\ref{fig:2}d-f). 
The wave center is shifted approximately 37 km northwestward of the airburst. It coincides with the location 
of meteorite tertiary disruption ($54.905^o N; 60.865^o E$) \cite{Popova_2013}. The analysis suggests 
that the TEC disturbances propagated radially up to 600-700 km and had wavelength around 220 km. 
The average velocity calculated from movement of front lines was $V=337\pm 89 m/s$. It should be 
noted that spherical wave is generally registered eastward, northeastward, and southeastward from 
the airburst. The interference of two wave types (spherical and cone-shaped) is observed westward, 
northwestward, and southwestward from the airburst. This effect is more expressed during 03:34-03:38 UT. 

\subsection{Coherent radar observations}

Measurements by GPS network are known to have difficulty in separating the spatial and temporal variation 
of ionospheric parameters. Unlike GPS data, SuperDARN  radars \cite{Chisham_et_al_2007} have high spatial 
and temporal resolution, and allows us to differ spatial and temporal effects. The first Russian coherent radar 
EKB is located $56.5^oN, 58.5^oE$ (near Ekaterinburg city), approximately 200 km to the North-West from the 
meteorite fall region  and operated with 60km spatial and 1min temporal resolution in the sector $ -6^o -  -44^o $ 
from North at distances up to 3500km. Fig.\ref{fig:1} shows the location of the radar and the orientation of each 
of the 16 rays (beams) of its field of view.

\begin{figure}
\includegraphics[scale=0.35]{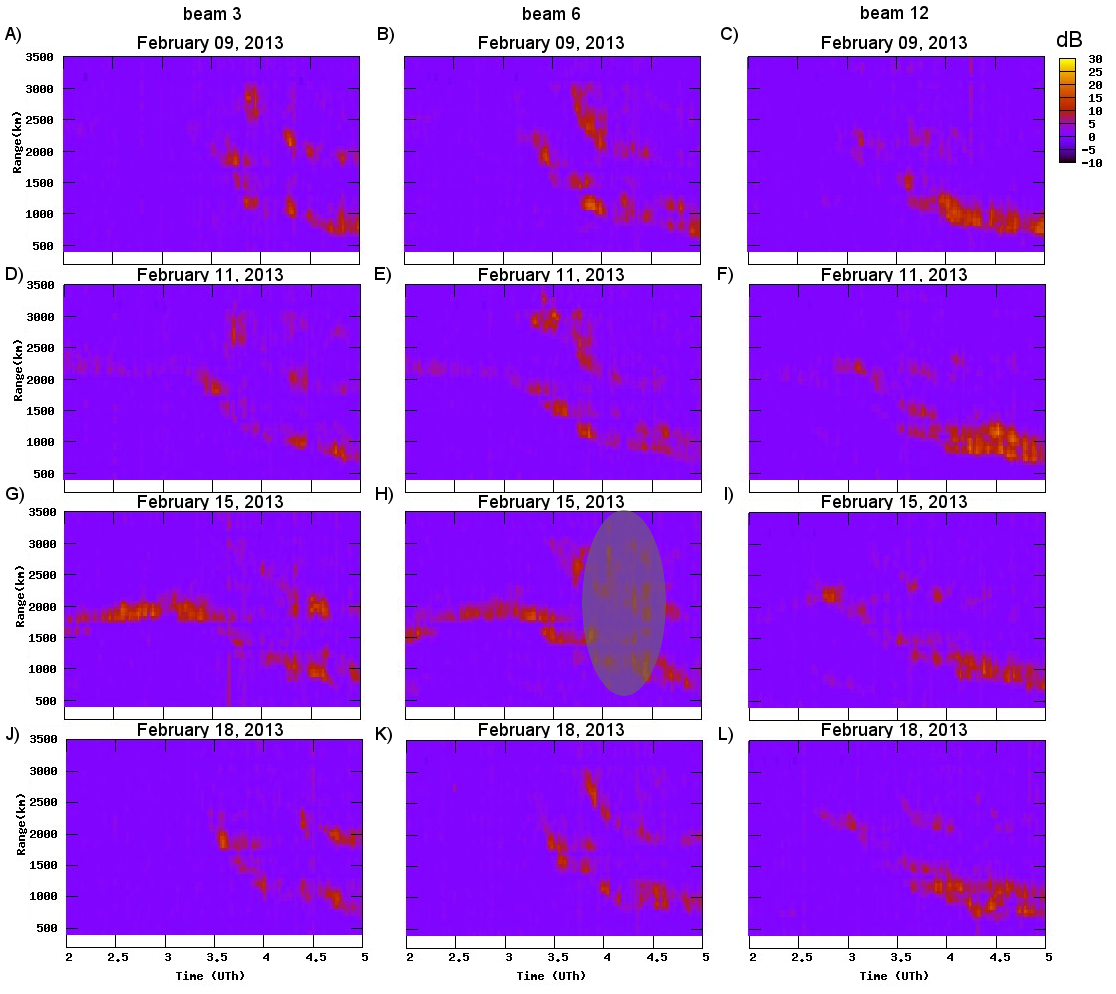}
\caption{
The intensity of the scattered signal during 02:00-05:00UT at 3,6,12 beams during 
geomagnetically quiet days February 09,11,15 and 18, 2013. Zone at H) marks the 
effect of refraction on MSTIDs arising after the flight and the explosion of the meteorite.
}
\label{fig:4} 
\end{figure}

After the meteorite fall, EKB radar recorded a large variety of ionospheric effects  \cite {Berngardt_2013a, 
Berngardt_2013b, Berngardt_2013c, Kutelev_2014}. The most powerful and basic effect is the observations of 
formation and radial passage of MSTIDs at E- and F-layer heights \cite{Kutelev_2014}. Fig.\ref{fig:4} presents 
the power of the scattered signal according to EKB radar data at 3,6,9,12 beams (Fig.\ref{fig:4}I-L) during 
February 15, 2013  02:00-05:00UT. In addition, the figure shows the power of the scattered signal in 
geomagnetically quiet days  - February 09, 11, 18, 2013.  The choice of these days been investigated 
in detail, for example, in \cite{Kutelev_2014}.

The tracks marked by areas 1-2 in the figure correspond to refraction of the radiosignal at ionospheric 
irregularities, which propagate radially from the radar with velocities 400 and 200 m/s at  E- and F-layer 
altitudes \cite{Kutelev_2014}.

According to Fig.\ref{fig:4}, the ionospheric effects in the field of view of the radar have strong azimuthal 
dependence. In particular, they present on 6 beam, and do not appear at 3 and 12 beams. To emphasize the 
azimuthal effects, we used the following technique:

- For each moment, for each beam, and for each range (this allows us to uniquely define each region in the 
field of view with approximately 100x100km size) we calculate the average power (in dB) higher than 
the noise level 0dB, over the magnetically quiet days February, 09,11,18, 2013; 

- Then, when processing 02/15/2013 data for each moment, beam and range we analyse only the data 
exceeding the threshold level (tripled the calculated average power at this point).

\begin{figure}
\includegraphics[scale=0.6]{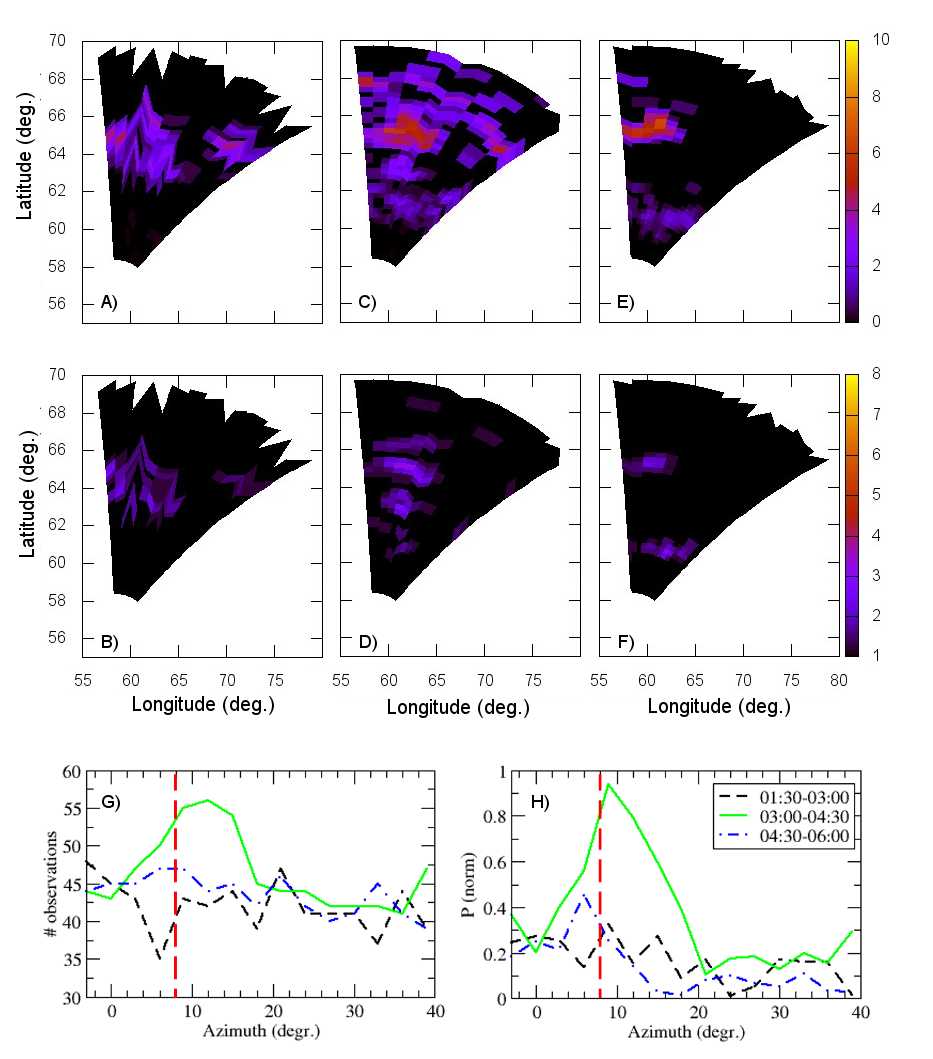}
\caption{
The normalized average scattering cross-section (A, C, E) and the number of ionospheric 
signals (B, D, F) in the field of view of the radar February 15, 2013, that exceed the 
threshold level for the appropriate geographic area, calculated over the geomagnetically 
quiet days. Estimates of MSTIDs are made for the periods: A-B) - 01:30-03:00 (before 
the meteorite fall); C-D) - 03:00-04:30 (during the 
meteorite fall); E-F) - 04:30-06:00 (after the meteorite fall); G-H) - total number of 
ionospheric signals (G), and the scattering cross section (H), as a function of sounding 
azimuth (from the north direction). The vertical dashed line is the perpendicular to the 
meteorite trajectory.
}
\label{fig:5} 
\end{figure}

- The power and the number of observations, accumulated for the given time period $[T_{start}, T_{stop}]$ 
for every fixed beam and every fixed range, are shown in Fig.\ref{fig:5}.

Fig.\ref{fig:5}A,C,E shows the intensity of the scattered signals in the radar field of view above the 
threshold level. Fig.\ref{fig:5}B,D,F presents the number of signals exceeding a threshold level.

As one can see from Fig.\ref{fig:5}C-D, the main effect is concentrated in the sector of directions 
$ 2^o - 18^o $ from the north. The dotted line marks the direction corresponding to the perpendicular 
to the meteorite trajectory.

\subsection{Seismic observations}

\begin{figure}
\includegraphics[scale=0.15]{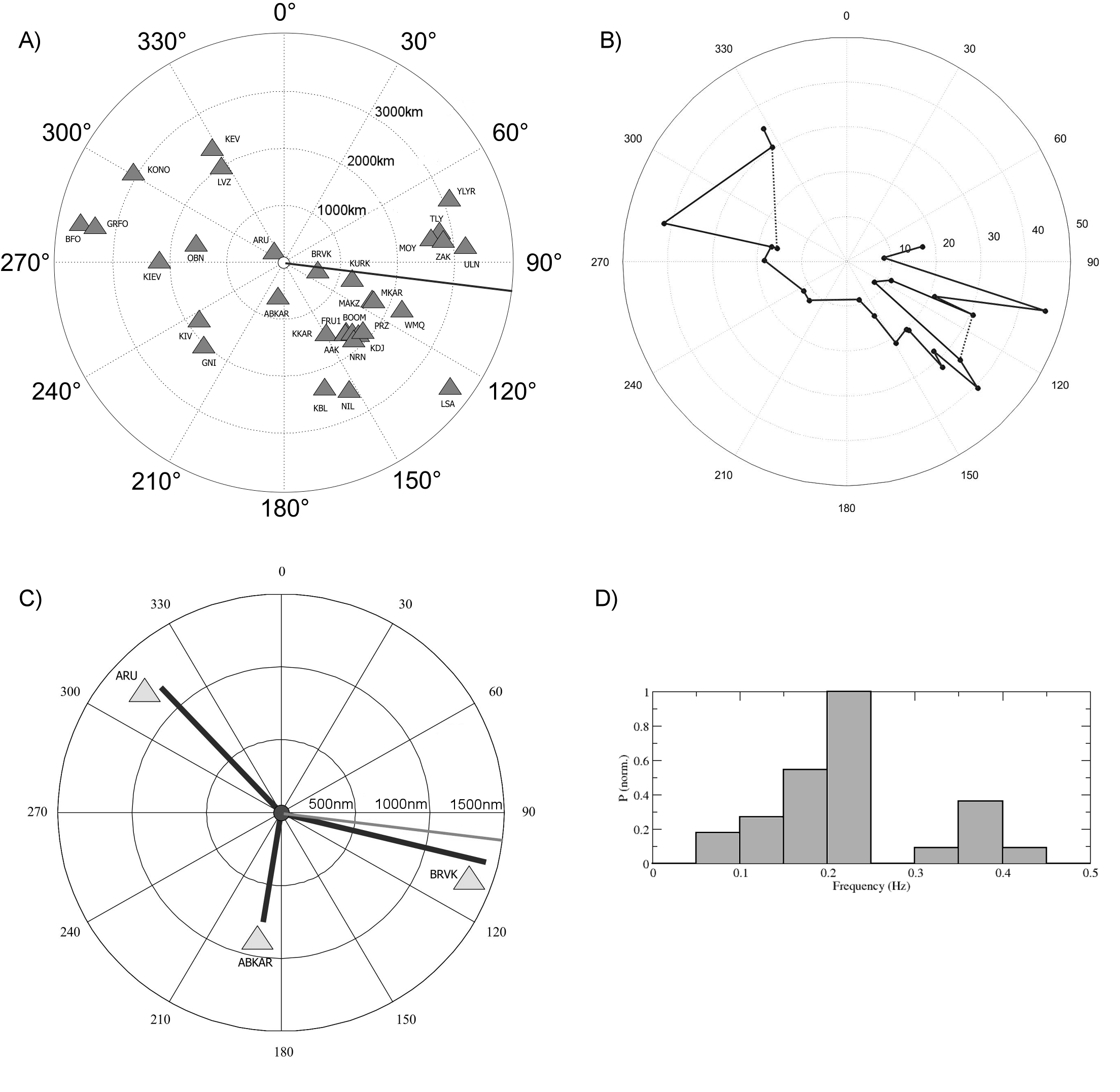}
\caption{
A) The azimuthal distribution of seismic stations (triangles) relative to the  point 
of the main bolide explosion (the center). The line marks the trajectory of the meteorite 
fall. B) The azimuthal distribution of vertical amplitudes reduced to epicentral 
distance 100 km. Dots show all the observed data (except data from ARU, BRVK, ABKAR 
stations  with epicentral distance <620 km). C) The azimuthal distribution of amplitudes 
of vertical seismic oscillations at stations ARU, BRVK, ABKAR (shown by triangles), gray 
line marks the trajectory of the meteorite fall, the center corresponds to the point of the 
main bolide explosion. D) The distribution of vertical oscillations frequency over the 
seismic stations.
}
\label{fig:10} 
\end{figure}

For the analysis of the seismic characteristics, we used seismograms obtained by worldwide networks of 
the broadband seismic stations Iris/Ida (II, http://ida.ucsd.edu/) and Iris/USGS (IU, http://earthquake.usgs.gov/regional/asl/) 
and by four regional networks: Iris/China (IC), Kazakhstan (KZ), Kyrgyzstan (KR) and Baikal regional 
seismological center (BY, Russia). Analysis of seismograms is a bit complicated due to presence of 
seismic waves from Tonga earthquake (February 15, 2013, 03:02:23UT, magnitude 5.8, 
coordinates 19.72S, 174.48W, depth 71.6 km). In the rest, the seismic situation was very quiet - no local or regional earthquakes  were registered on 
the seismograms. Visual analysis revealed the presence 
of the surface seismic waves at some stations that may be associated with meteorite according to their 
arrival times. These waves were recorded at 32 seismic stations (Fig.\ref{fig:10}A), located at distances 
$r_0$ from 252 to 3654 km from the epicenter, and they are short-period oscillations with a period 
T = 3-16 sec and with duration up to 1 minute. For stations situated more than 3600 km from the fall 
site, a surface wave is lost in the seismic waves from the Tonga earthquake. The group velocity $C_r$ 
of Rayleigh waves for periods of 20-100 seconds varies from 2.3 to 3.5 km/s. 

For the analysis of azimuthal distribution of surface seismic waves the maximal amplitudes of vertical 
oscillations $U_z^{R}$ (Fig.\ref{fig:10}B), and the spectral corner frequencies $f_C$  were analyzed 
(Fig.\ref{fig:10}D). The distribution of seismic stations around the point of the explosion is non-uniform: 
there is a very big azimuthal gap approximately from $330^o$ to $70^o$. For this azimuthal range it is impossible 
to allocate the surface wave from the records due to the superposition of waves from the Tonga earthquake, as 
well as to the lack of broadband seismic stations ( Fig.\ref{fig:10}A). Analysis of $U_z^{R}$  at the nearest 
stations ($r_0 <$  620 km) shows that the maximum $U_z^{R}$ = 1414.4 nm is observed at station BRVK 
($r_0$= 616 km, azimuth - $104^o$), at the same time, the amplitude at station closer to the fall point is 
smaller: at ARU station ($r_0$ = 252 km, azimuth $317^o$) $U_z^{R}$ =1177.5 nm, at ABKAR station 
( $r_0$ = 618 km, azimuth $189^o$) - $U_z^{R}$ = 762.4 nm (Fig.\ref{fig:10}C).

The azimuthal distribution of the maximum amplitudes of the Rayleigh waves at different stations indicates 
the absence of spherical symmetry (Fig.\ref{fig:10}B) - the maximum amplitudes are oriented along the 
direction of the meteorite trajectory, the minimum amplitudes are observed in perpendicular directions. 
There is three exceptions: an abnormally low amplitudes are observed at LSA (distance $r_0$ = 3654 km) 
and MKAR ($r_0$ = 1709 km) stations, located in azimuths $113^o-127^o$, and abnormally high amplitude 
at the KONO station (azimuth $301^o$, $r_0$ = 3073 km). But statistically, the azimuthal distribution 
of the Rayleigh waves amplitudes $U_z^{R}$ has two peaks elongated with the meteorite trajectory: the 
first maximum is oriented towards the meteorite fall direction and the second - in the opposite direction. 

The analysis of the azimuthal distribution of $U_z^{R}$ showed that the distribution elongated with the 
meteorite fall trajectory (Az = $97^o$). At the same time, we found no spherical symmetry in 
seismic wave radiation that corresponds to bolide explosions. This allows to suggest that the seismic effect is caused by single extended event rather than a series of explosions. This fits well in with 
the model \cite{Popova_2013}. An explanation of the observed asymmetry in the seismic wave amplitudes 
from the single point source is also not satisfactory. In some cases, azimuthal nonuniformity of the 
seismic wave amplitudes may be explained by an anisotropy of medium. But this explanation does not 
work in our case, as the main tectonic structures in the Ural region are oriented orthogonally to 
the meteorite trajectory. Thus, we can conclude that there is a directivity of seismic radiation. 

The effect of directivity is reflected both in the seismic wave amplitudes and on the corner frequencies. 
The concept of the corner frequency is associated with the rupture parameters (its velocity and length) 
and directivity function \cite{BenMenahem_1961}: 

\begin{equation}
D(\omega)=\frac{C_{r}/V+cos\theta}{C_{r}/V - cos\theta}
\cdot
\frac{sin \left( \frac{\omega b}{2C_{r}} \left( C_{r}/V - cos\theta \right) \right)} {sin \left( \frac{\omega b}{2C_{r}} \left( C_{r}/V + cos\theta \right) \right) }
\end{equation}

where $C_r$ is the seismic wave velocity, $V$ is the rupture speed, $b$ is the length of rupture, 
$\omega$ is angular frequency, $\theta$ is the angle  between the rupture propagation direction and 
the direction to the seismic station. 

This function has a series of maxima and minima at frequencies, their positions depend on the 
values $b$ and $V$. The first maximum/minimum corresponds to a frequency $f_C=\omega/(2\pi)$ 
called the corner frequency. Thus, the corner frequency $f_C$ includes the effect of the propagation 
of rupture and it may be measured directly from the surface wave spectrum. It corresponds to the 
frequency of the  inflection point of the seismic spectrum, separating the horizontal plateau at 
low frequencies and decaying oscillations with sloping envelope at high frequencies. For the 
meteorite surface waves, the corner frequencies vary widely (0.06-0.43 Hz), but the maximum 
number of stations observe frequencies 0.15 - 0.25 Hz (Fig.\ref{fig:10}D).

\section{Models}

\subsection{Acoustic signal model}

To explain the azimuthal characteristics of ionospheric effects, we made mathematical simulations of acoustic signal propagation at different heights.

Models based on  one or more explosions are often considered as model caused by the ionospheric 
disturbances. However, from our point of view, the model of 
ballistic  wave described in (SM in \cite{Popova_2013}, pp.62-69) is the most realistic one.
It successfully explains the observed 
destructions (which are associated with a passage of powerful acoustic wave) as continuous formation 
of perturbations along the meteorite trajectory. Similar models have been successfully used for 
explaining the zone of destruction after the Tunguska meteorite fall \cite{Zotkin_1966, BenMenahem_1975}, 
which indicate a high degree of this model reliability. In this model, the amplitude of the source along 
the trajectory is determined by the luminosity of the bolide \cite{Zotkin_1966}. The disadvantages of 
using other models to describe the destruction zone in the case of 'Chelyabinsk' meteorite  were 
investigated in detail in (SM in \cite{Popova_2013},pp.62-69).

Therefore, we used the simple model of the summation of partial pressures coming from each point 
of the meteorite trajectory with taking into account the acoustic propagation delay. We used the 
model of a moving source. The initial acoustic pressure disturbance is proportional to the 
luminosity of the bolide at the meteorite position for given moment \cite{Zotkin_1966, BenMenahem_1975}. 
The luminosity is shown in  (SM in \cite{Popova_2013}, pp26-27). 

As a result of interference, the meteorite fight forms the focus of the acoustic wave - Mach cone (ballistic 
wave). In a homogeneous medium for the meteorite velocity  17.5km/s this wave spreads nearly perpendicularly 
to the flight trajectory. In a medium with nonuniform acoustic speed height profile the Mach cone shape 
substantially transforms. This transformation is determined by many factors, including the acoustic speed 
at different heights. Therefore at different distances, the angle between propagation direction and the 
meteorite trajectory becomes different.

For estimates, we used the qualitative model for dependence of the acoustic wave pressure on the neutral 
density and acoustic speed \cite{Yuen_1969}:

\begin{equation}
P(\overrightarrow{r},t)=\sqrt{\frac{\rho(\overrightarrow{r})C_{s}(\overrightarrow{r})}{\rho(\overrightarrow{R_{0}})C_{s}(\overrightarrow{R_{0}})}} \\
P_s(\overrightarrow{R_{o}},t-T_{s}(\overrightarrow{R_{0}},\overrightarrow{r}))
\label{eq:1}
\end{equation}

Here, the delay $T_s$ is determined by trajectory of the acoustic signal and based on the 
geometrical optics laws for the acoustical signal in inhomogeneous media \cite{Blokhintsev_1952, Clay_1977}, 
where the refractive index is determined by the acoustic speed $ C_s (\overrightarrow{R}) $. 
The $ \rho (\overrightarrow{R}) $  is the density of the neutral atmosphere and 
$ P_s (\overrightarrow{R_{0}}, t) = \alpha I (\overrightarrow {R_{o}}, t) $ is the initial acoustic pressure 
of the source, which (in the first approximation) is equal to the luminosity of the bolide 
$I(\overrightarrow {R_{o}}, t) $ 
at this point at corresponding moment \cite{Zotkin_1966, BenMenahem_1975} with accuracy of 
a constant multiplier $\alpha$. $\overrightarrow{R_{0}}=\overrightarrow{R_{0}}(t)$ is the 
position of the bolide in the moment $t$.

The resulting partial acoustic pressure $P_s$ at each point of $(\overrightarrow{r}, t) $ is 
summed up with taking into account the geometroacoustic propagation delay $ T_s $ from the 
bolide (source) point to the  specified point. Since the source of the acoustic wave at each 
moment is assumed isotropic, we conducted an additional integration over all the acoustic 
signal propagation directions from the source point.

The duration of the meteorite fall in the upper atmosphere was $ \approx 16 sec $, 
which corresponds to a wavelength $ \approx 5km $, considerably less than a hundred 
kilometers - the distance to the studied effects. This can qualitatively substantiate the 
applicability of geometrical acoustics for qualitative estimates of the observed ionospheric 
effects. In the calculations of the acoustic field, we assumed that the acoustic speed and the 
neutral density profile are functions of the height $ H $. The earth surface is considered to be flat, 
due to the smallness of the investigated distances. The acoustic speed and neutral density profile 
were estimated from the reference model NRLMSIS-00 for observation point.

As an initial pressure perturbation $P_s(\overrightarrow{R_0}, t)$, forming an acoustic signal 
we used a short pulse propagating at the velocity $ V_{0} = 17.5km/s $ along the meteorite  
trajectory from 90km to 25km altitude.  $ P_s(\overrightarrow{R_0},t) $ depends on the height 
and luminosity curve defined in (SM in \cite{Popova_2013}, pp26-27). We used the data given in 
\cite{Borovicka_2013} for entry point $ r_{0} $, the trajectory and meteorite fall velocity $ V_{0} $.

It should be noted that the model is very rough, and does not include neither details of the acoustic 
signal amplitude changes with height, nor conversion of neutral density variations into electron density 
variations. The effects also depends on the height \cite{Maruyama_2014} and their presence is required 
for a correct estimation of the amplitude of ionospheric effects. However, as it will be shown below, the 
model is suitable for qualitative explanation of azimuthal and temporal effects.

\begin{figure}
\includegraphics[scale=0.54]{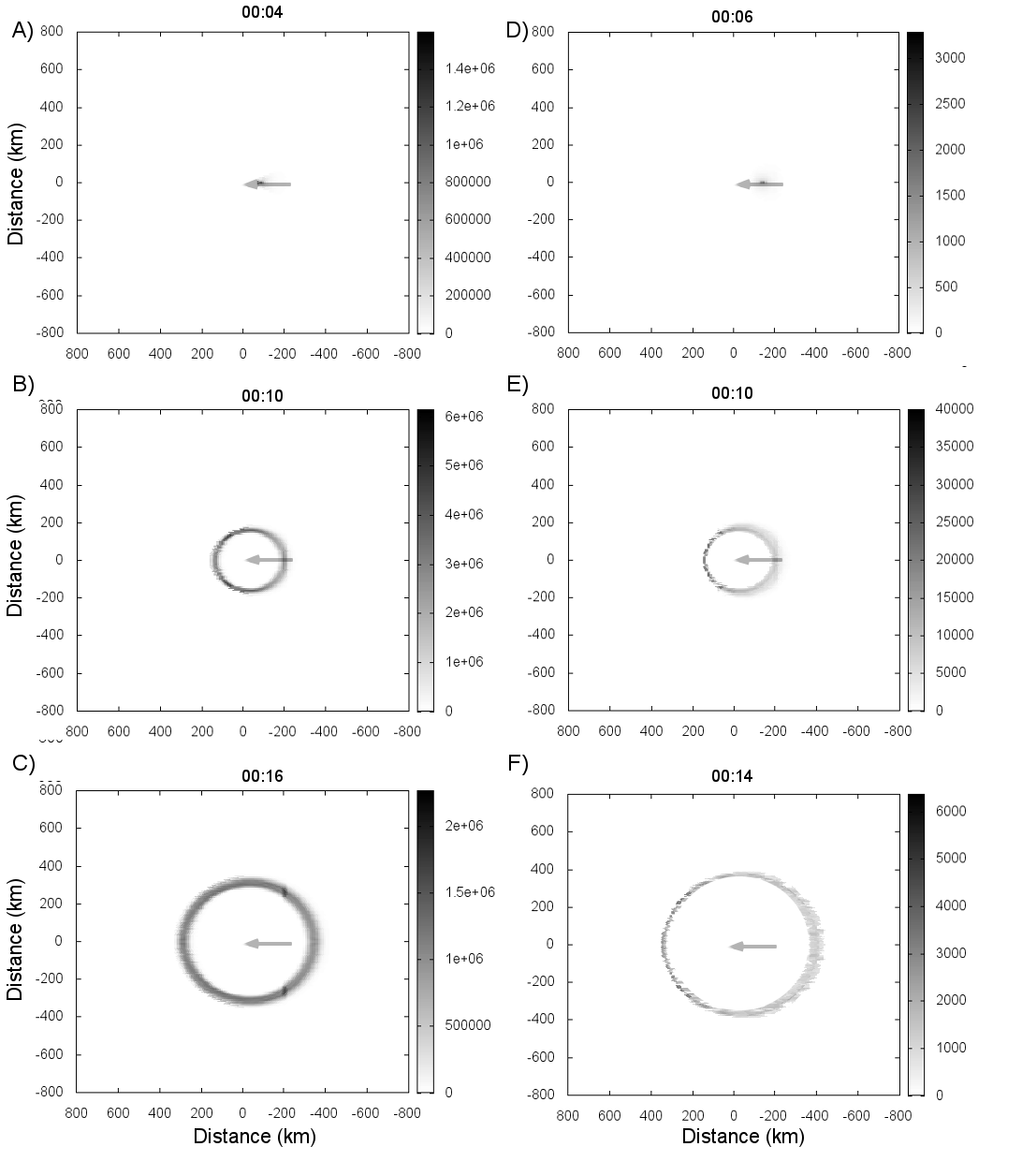}
\caption{
The dynamics of the acoustic wave pressure at 120km (left) and 240 km (right). Arrow 
marks the meteorite trajectory.
}
\label{fig:11b} 
\end{figure}

The evolution of amplitude at 120km altitude, in E-layer, as a function of time, is shown at Fig.\ref{fig:11b}a-c. 
One can see from the figure that the main acoustic effects should be observed 4-6 minutes after the explosion and 
spread further at the acoustic speed ($ \approx 400m/s $) in a direction perpendicular to the meteorite fall 
trajectory.

Fig.\ref {fig:11b}d-f presents the evolution of acoustic wave pressure at 240km, in the F-layer, as a function 
of time. It shows that the main acoustic effects should be observed 7-8 minutes after the explosion and should 
propagate with acoustic speed ($ \approx 800m/s $) in a direction of $40^o-50^o$ to the meteorite trajectory.

\begin{figure}
\includegraphics[scale=0.48]{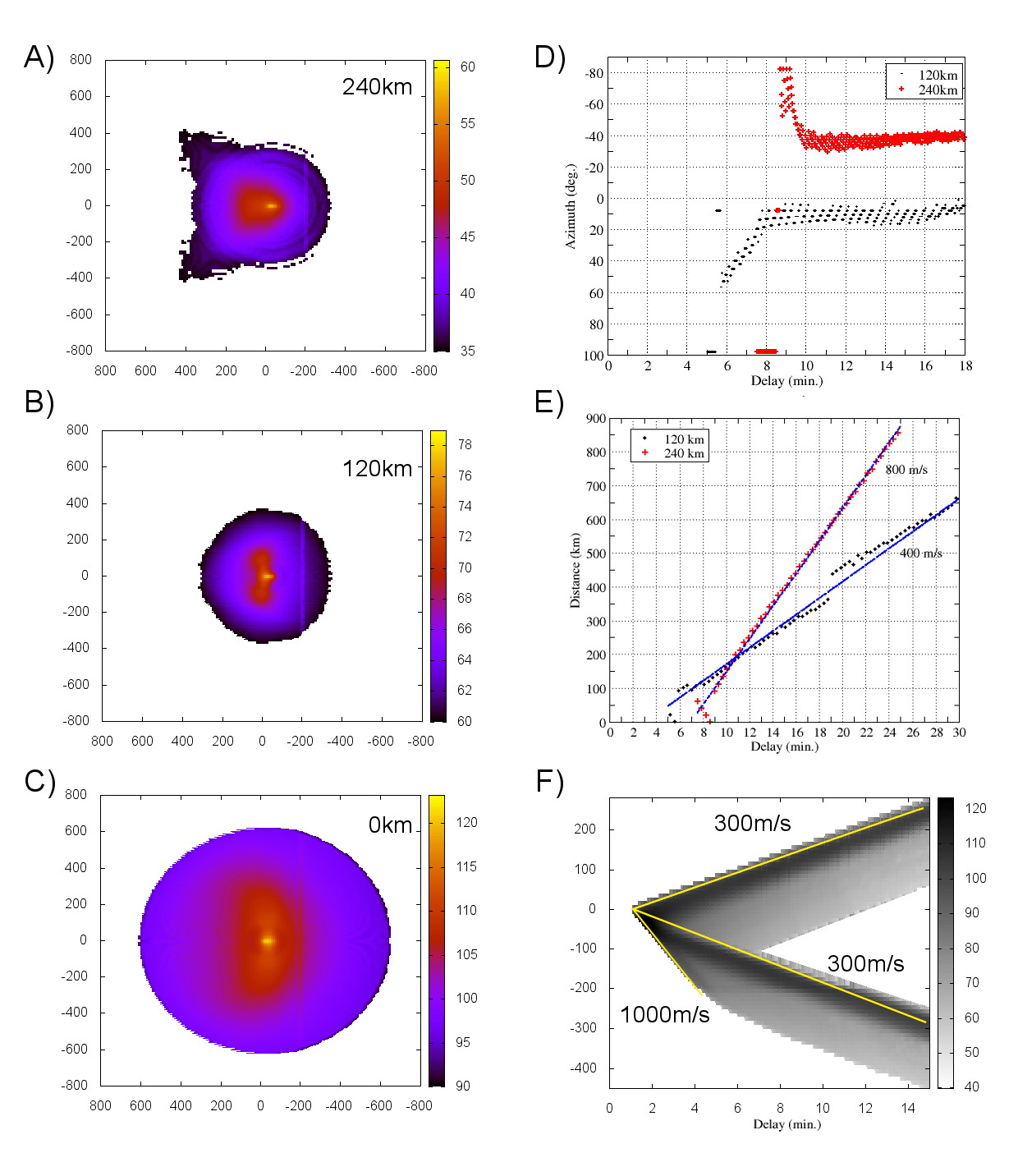}
\caption{
Model of acoustic wave pressure at different heights: A) - 240km; B) - 120km; C) - 0km; 
D) - azimuth (from the north) of ballistic wave propagation at 240 km and 120 km heights; 
E) - velocity of ballistic wave propagation at 240 km and 120 km  heights; F) - 
propagation velocity of ballistic wave maximal intensity on the Earth surface.
}
\label{fig:11c} 
\end{figure}

Fig.\ref{fig:11c}a-c shows the modeled acoustic wave maximal pressure at various heights and at various 
distances from the main bolide explosion site, with taking into account the acoustic sound speed profile 
at the upper atmosphere. 

It can be seen that the acoustic wave pressure at different altitudes has a strong  azimuthal dependence. 
At relatively low altitudes (below 150 km),  the highest acoustic wave pressure is predicted at azimuths 
perpendicular to the meteorite trajectory. This fits well with the results of \cite{Popova_2013}  which 
associate the azimuthal features of the destruction zone on the earth's surface with an acoustic wave pressure. 
This justifies the possibility of using the simple model (\ref{eq:1}) for estimation of azimuthal effects. At 
altitudes above 150 km, the model predicts an increase of the acoustic wave pressure at azimuths $\approx 45^o $ 
to the meteorite trajectory in the direction of the meteorite fall.

From Fig.\ref{fig:11c}A-C one can see that the highest acoustic wave pressure is observed under the trajectory 
of the meteorite. Fig.\ref{fig:11c}C shows the dynamics of the acoustic wave pressure along the meteorite 
trajectory projection to the Earth's surface. As it is shown at Fig.\ref {fig:11c} F,  
three acoustic disturbances propagate along the surface of the Earth: the most powerful propagates 
back towards its fall with {$ \approx 1 km/s$} velocity, which is defined by the meteorite trajectory fall 
angle, the meteorite speed and the acoustic speed profile. This velocity roughly coincides with the estimates 
given in  (SM in \cite{Popova_2013}, p.70), which also confirms the possibility of using our model to estimate 
the effects. The two other waves are acoustic ones, having velocity  $\approx 320m/s$, which is associated 
with the propagation of spherical acoustic wave at ground altitudes.

\subsection{Seismic signal model}

The propagation velocity of surface seismic waves was about 3-3.5km/s \cite{Tauzin_2013}. Therefore 
the seismic oscillations at longer distances can not be directly caused by atmospheric acoustic waves 
having the velocity 10 times smaller. However, the movement of a supersonic powerful acoustic pulse 
(ballistic wave) on the ground can cause the observed seismic effect indirectly. This model, for example, 
was used for an explanation of seismic effects of Tunguska meteorite \cite{BenMenahem_1975}. To estimate 
the characteristics of the effect we carried out numerical calculations of the signal intensity in a simple 
model developed for the analysis of seismic signals generated during formation of geological faults. As the 
fault line we use the trajectory of the acoustic signal on the ground. Parameters of fault formation (length, 
velocity and direction) are equal to the parameters of the acoustic pulse movement on the ground.

To calculate the theoretical seismic radiation pattern we used the model proposed in \cite{BenMenahem_1961} 
for a linear fault. According to this approach, the azimuthal distribution of the seismic waves is defined by 
the type of the fault and by the parameters of its formation: the rupture speed $V$, its length 
$b$, the frequency of seismic waves $ \omega $, and the phase velocity of Rayleigh waves $C_r$.

In our case, two types of ruptures with the vertical position 
of the plane displacement may be considered as a seismic analogue - vertical strike-slip fault and 
the vertical dip-slip fault. In both cases, the maximal amplitudes have to orient according to 
the slip vector: for the strike-slip fault – they are oriented in the direction strike fault, and 
for dip-slip fault – they are perpendicular to the fault plane \cite{Kasahara_1981, Udias_2014}.

In our analysis, we used the following formulas that qualitatively describe (up to a constant multiplier) the 
azimuthal distribution of amplitudes of vertical seismic oscillations $ U_ {z} ^ {R} $ in cases of different  
fault types. Thus, the amplitude of the seismic waves depends strongly on the angle $ \theta $ between fault 
propagation direction and the direction to the seismic station. The amplitude also depends on the distance to 
the observation point $r_0$ .

According to \cite{BenMenahem_1961}, in the case of vertical strike-slip fault the azimuthal distribution of 
the Rayleigh displacement for the vertical component is given by:

\begin{equation}
 U_{z}^{R}=\frac{cos \theta }{\sqrt{r_0}} \frac{sin X_r}{X_r}
 \label{eq:5}
\end{equation}

where 
\begin{equation}
 X_r=\frac{\omega b}{2 C_r} \left( \frac{C_r}{V} - cos \theta \right)
 \label{eq:6}
\end{equation}

In the case of vertical dip-slip fault the azimuthal distribution of the Rayleigh displacement 
for the vertical component is given by:

\begin{equation}
 U_{z}^{R}=\frac{1}{\sqrt{r_0}} \frac{sin Y_r}{Y_r}
 \label{eq:7}
\end{equation}

where 
\begin{equation}
 Y_r=\frac{\omega b cos \theta}{2 C_r} 
 \label{eq:8}
\end{equation}

For the simulation we choose the propagation line of the fault as the trajectory of the meteorite 
fall (azimuth of propagation $ 97^o $). 

We suppose the rupture speed $ V \approx 1km/s $, its length 
$ b \approx 40km $, the frequency of seismic waves $ \omega $, defined from fig.\ref{fig:10}D and the phase velocity of Rayleigh waves $ C_r \approx 3.4km/s $.

The direction of the rupture was opposite to the direction 
of the fall. This corresponds to the direction and the trajectory of the ballistic wave peak on 
the ground.

~

\section{Discussion}

\begin{figure}
\includegraphics[scale=0.15]{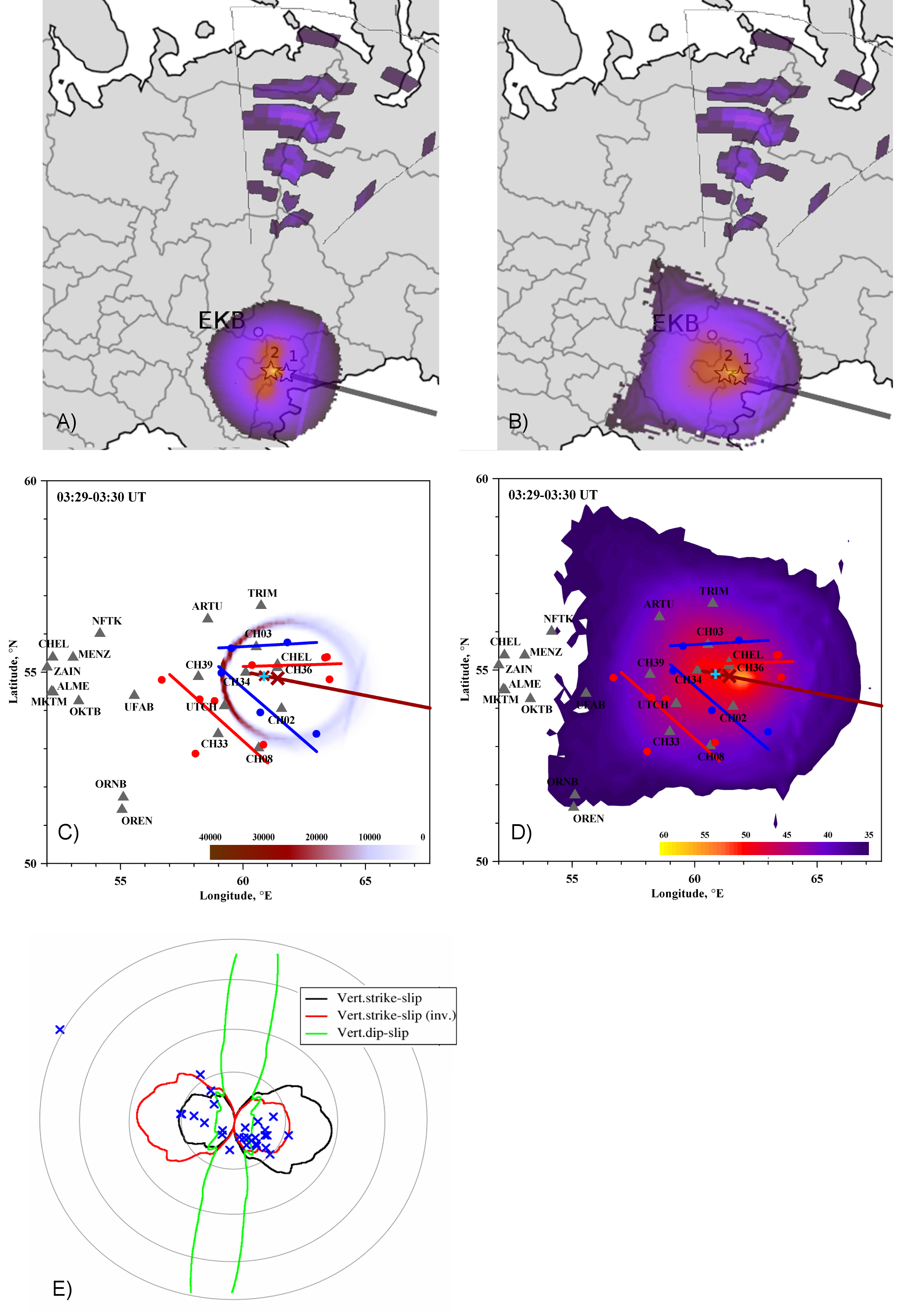}
\caption{
A-B): Comparison of scattered signal power variations that are observed by EKB radar (A) 
and the model estimates of the acoustic wave pressure at 120 km height (B); C-D): TEC 
disturbances and distribution of acoustic wave pressure.  Blue dots correspond to 
minimums in $\Delta TEC$, red ones correspond to maxima in $\Delta TEC$. Lines mark 
approximations of the wavefront over the data. Additionally shown C) - a model distribution 
of the acoustic wave pressure at 03:30UT at 240 km height, D) - a model of the acoustic wave 
pressure for the period of 03:20-03:50UT at 240 km.
E) The azimuthal distribution of the amplitudes of  vertical seismic oscillations, multiplied 
by $ \sqrt {r_0} $ (blue crosses) and the model azimuthal distribution of amplitudes of 
vertical seismic oscillations (\ref {eq:5}, \ref {eq:7}) (lines). The expected distribution 
in the case of vertical dip-slip fault is shown by green line, black and red lines show the 
azimuthal distribution of the amplitudes for the vertical strike-slip fault. The black and red 
lines correspond to the cases of opposite and direct fault propagation relatively to the 
meteorite fall direction.
}
\label{fig:12} 
\end{figure}

Fig.\ref{fig:12}A,B presents a comparison between observed MSTIDs and acoustic wave pressure 
distribution at 120km. One can see that the model of acoustic waves in the lower ionosphere 
describes the experimental data sufficiently well. The same conclusion can be made from a 
comparison of Fig.\ref{fig:11c}c and Fig.\ref{fig:5}. According to the estimates carried out before, 
the velocity of azimuth-dependent ionospheric irregularities was about 400m/s \cite{Kutelev_2014}. 
This corresponds well to the velocity of acoustic wave pressure variations at altitudes 120-140km. 
So we can conclude that the MSTIDs, observed by EKB radar, are well described by the effect of the ballistic 
wave passage at altitudes of 120-140km. This also stimulates similar disturbances in the charged 
components of ionospheric plasma (for example, by mechanism described in \cite{Maruyama_2014}).

As simulations show, the acoustic radiation pattern at 240km has a specific two-horned 
structure, which describes the experimental results observed in the first 14 minutes after 
the explosion sufficiently well.  Fig.\ref{fig:12}C-D shows the imposition of model calculations 
and experimental observations. We can see from Fig.\ref{fig:12}C that not only irregularities 
azimuthal distribution, but also the expected moment of  irregularities observation are in a 
good agreement with the model. Estimates of the propagation velocity of TEC wave cone 6-14min 
after explosion is $661 m/s \pm 256m/s $, and are in a good agreement with the calculated acoustic 
wave pressure velocity at 240km height (Fig.\ref{fig:11c} E). The observation of such waves 
by the EKB radar is hampered by the fact that the azimuthal effect is located outside the 
antenna pattern.

14 minutes after the explosion we observed a formation and propagation of a spherical MSTID 
with velocity $ 337 \pm 89 m/s $. Similar midscale (wavelength of 200-300 km) concentric 
MSTIDs with velocities $ 138-423 m/s $ were detected by GPS receivers after the Tohoku 
earthquake (March 11, 2011) \cite{Tsugawa_2011}. The earthquake-generated MSTIDs began 
to be observed at distances $\approx 300 km$ from the epicenter, and propagated up to 1000km. 
According to the numerical simulation \cite{Matsumura_2011}, these propagating spherical 
MSTIDs  can be related with gravitational mode of internal atmospheric waves.

Thus, by GPS network we observed the two MSTID kinds in TEC: the conical wavefronts associated 
with ballistic wave 6-14 minutes after the explosion, and the spherical wavefronts associated 
with internal airwave 14 minutes after the explosion.

Fig.\ref {fig:12}E shows the azimuthal distribution of the amplitude of vertical seismic 
oscillations, multiplied by $ \sqrt {r_0} $ and the model azimuthal distributions 
(\ref{eq:5}, \ref{eq:7}). The model azimuthal distribution was calculated as the average 
of azimuthal distributions of expected vertical seismic oscillations, with taking into 
account the experimental distribution of their measured frequency (see Fig.\ref {fig:10} D). 
Azimuthal distribution of normalized amplitudes, observed experimentally, is shown with 
blue crosses, the expected model azimuthal distributions are shown by lines. Green line 
corresponds to the case of vertical dip-slip fault, black and red lines correspond to the 
case of the vertical strike-slip fault.  The black and red correspond to the cases of fault 
propagation in opposite and direct directions of the meteorite fall respectively.

The figure shows that the model of vertical strike-slip fault describes qualitatively 
the observed azimuthal distribution of the amplitudes of the vertical oscillations, and 
its elongation with the trajectory of the meteorite. This allows us to interpret ballistic 
wave peak, moving by the ground with 1km/s velocity, as an equivalent vertical strike-slip 
fault, generating seismic waves, propagating with Rayleigh velocity.

\section{Conclusion}

We present the results of a studying the azimuthal characteristics of ionospheric and 
seismic effects of the meteorite 'Chelyabinsk' based on the data from the GPS receivers 
network, coherent decameter radar EKB SuperDARN and network of seismic stations located 
near the meteorite fall trajectory. 

It is shown that 6-14 minutes after the bolide explosion network of GPS receivers steadily 
observed the cone-shaped wavefront of MSTIDs that is interpreted as a ballistic acoustic 
wave. We assume the 240km effective ionospheric height. The typical MSTIDs propagation 
velocity were observed $661 \pm 256 m/s $, which corresponds to the expected acoustic 
wave speed for 240km height.

14 minutes after the bolide explosion, at 200km distance we observed the emergence and 
propagation of a MSTID with spherical wavefront, that may be interpreted as gravitational 
mode of internal acoustic waves. The propagation velocity of this MSTID were $ 337 \pm 89 m/s $ 
which corresponds to the propagation velocity of these waves in similar situations. 

At EKB SuperDARN radar, we observed MSTIDs in the sector of azimuthal angles close 
to the perpendicular to the meteorite trajectory. The observed MSTID velocity (400 m/s) 
and azimuthal properties correlate well with the model of ballistic wave propagating 
at 120-140km.

Thus, our modeling shows that ballistic wave produced by supersonic motion of the 
meteorite, and its burning in the upper atmosphere can satisfactorily explain the 
various azimuthal ionospheric effects occurring during the first minutes after the explosion. 
The azimuthal dependence of ionospheric effects is observed by the coherent EKB SuperDARN 
radar and GPS network. 14 minutes after the explosion in the ionosphere additional spherically 
symmetric perturbations are beginning to occur that are associated with generation of 
gravitational mode of internal atmospheric  waves.

It is shown that the azimuthal dependence of the amplitude of vertical seismic 
oscillations can be described qualitatively by the model of vertical strike-slip 
fault, propagating at the velocity 1km/s along the trajectory of the meteorite to 
the distance of about 40km. These parameters correspond to the velocity of ballistic 
wave pulse propagation by the earth surface. This allows us to consider the ballistic 
wave to be a moving source of seismic oscillations for a qualitative description of 
seismic observations during the meteorite fall.

Our modeling and experimental data show that ballistic wave, caused by the 
supersonic flight of the bolide, plays a decisive role for the observed 
azimuthal dependence of ionospheric and seismic effects.


%
%
%
%
%
%
%

\section*{acknowledgments}
We are deeply indebted to LLC "GEOSalyut" (Moscow) and personally to S. Parshin 
as well as to LLC "Poleos" (Chelyabinsk) for access to data from Chelyabinsk GPS 
network, to NAVGEOKOM company (http://www.navgeocom.ru) for access to data from 
the Russian GPS network, to Scripps Orbit and Permanent Array Center (SOPAC, 
http://sopac.ucsd.edu) for access to data from the global GPS network.  
The facilities of the IRIS Data Management System, and specifically the IRIS 
Data Management Center, and Baikal Regional Seismological Center were used for 
access to waveform and metadata required in this study.
The work of G.A.Zherbtsov  was supported by the RF President Grant of Public 
Support for RF Leading Scientific Schools (NSh-2942.2014.5). The work of 
O.I.Berngardt, N.P.Perevalova, A.A.Dobrynina, K.A.Kutelev, O.A.Kusonsky 
and N.V.Shevtsov was supported by RFBR grant 14-05-00514. The work 
of N.V.Shevtsov was also supported by the Far Eastern Federal University, 
project No. 14-08-01-05\_m.

\newpage

\begin{thebibliography}{1}
\bibitem[{{Afraimovich et al.}}{1998}]{Afraimovich_1998} Afraimovich, E. L., K. S. Palamartchouk, and N. P. Perevalova (1998), GPS radio interferometry of travelling ionospheric disturbances// \textit{J. Atmos. Solar-Terr. Phys.}, \textit{60}, 1205-1223. 
\bibitem[{{Afraimovich et al.}}{2009}]{Afraimovich_2009}Afraimovich, E. L., I. K. Edemskiy, A. S. Leonovich, L. A. Leonovich, S. V. Voeykov, and Yu. V. Yasyukevich (2009), The MHD nature of night-time MSTIDs excited by the solar terminator// Geophys. Res. Lett., 36, L15106, doi:10.1029/2009GL039803.36. 
\bibitem[{{Alpatov et al.}}{2013}]{Alpatov_2013}Alpatov, V.V., D. V. Davydenko, V. B. Lapshin, E. S. Perminova, V. A. Burov, L. N. Leshenko, J. I. Portnyagin, G. F. Tulinov, J. P. Vagin, D. S. Zubachyov, J. S. Rusakov, M. A. Chichaeva, V. N. Ivanov, D. A. Lysenko, K. A. Galkin, V. T. Minligareyev, N. L. Stal, V. S. Chudnovsky, A. N. Karhov, A. V. Syroeshkin, A. Y. Shtyrkov, Y. V. Gluhov, V. A. Korshunov, M. A. Morozova, and A. V. Tertyshnikov (2013), Geophysical conditions at the explosion of the Chelyabinsk (Chebarkulsky) meteoroid in February 15, 2013// Heliogeophysical Research, Moscow, 37 p. http://vestnik.geospace.ru/index.php?id=180 (in Russian). 
\bibitem[{{Ben-Menahem}}{1961}]{BenMenahem_1961} Ben-Menahem A., Radiation of seismic surface waves from finite moving sources // Bull. Seism. Soc. Am. 1961. V. 51. P. 401-435. 
\bibitem[{{Ben-Menahem}}{1975}]{BenMenahem_1975}Ben-Menahem A., Source parameters of the Siberia explosion of June 30, 1908, from analysis and synthesis of seismic signals at four stations// Physics of the Earth and Planetary Interiors, 11, 1-35, 1975.
\bibitem[{{Berngardt et al.}}{2013a}]{Berngardt_2013a}Berngardt, O. I., A. A. Dobrynina, G. A. Zherebtsov, A. V. Mikhalev, N. P. Perevalova, K. G. Ratovskii, R. A. Rakhmatulin, V. A. San’kov, and A. G. Sorokin (2013a) Geophysical Phenomena Accompanying the Chelyabinsk Meteoroid Impact// Doklady Earth Sciences, 452(1), 945-947. 
\bibitem[{{Berngardt et al.}}{2013b}]{Berngardt_2013b}Berngardt, O.I., V. I. Kurkin, G. A. Zherebtsov, O. A. Kusonski, and S. A. Grigorieva (2013b), Ionospheric effects during first 2 hours after the Chelyabinsk meteorite impact// arXiv:1308.3918 physics.geo-ph, 30 p. 
\bibitem[{{Berngardt}}{2013c}]{Berngardt_2013c}Berngardt, O. I. (2013c), Seismo-ionospheric effects associated with "Chelyabinsk" meteorite during the first 25 minutes after its fall// arXiv:1409.5927 physics.geo-ph, 22 p. 
\bibitem[{{Blokhintsev}}{1952}]{Blokhintsev_1952} Blokhintsev, D. I., The Acoustics of an Inhomogeneous Moving  Medium (trans. from russian R. T. Beyer and D. Mintzer). Providence: Brown University  Research Analysis Group,  1952.
\bibitem[{{Borovicka et al.}}{2013}]{Borovicka_2013} Borovicka J., Pavel Spurny P. and Shrbeny L., Electronic Telegram No. 3423 // Central Bureau for Astronomical Telegrams, International Astronomical Union, 2013 February 23 
\bibitem[{{Chernogor}}{2015}]{Chernogor_2015}Chernogor L.F., Ionospheric effects of the Chelyabinsk meteoroid// Geomagnetism and Aeronomy, 55(3), pp 353-368, 2015
\bibitem[{{Chisham et al}}{2007}]{Chisham_et_al_2007}Chisham G. , M. Lester, S. E. Milan, M. P. Freeman, W. A. Bristow, A. Grocott, K. A. McWilliams, J. M. Ruohoniemi, T. K. Yeoman, P. L. Dyson, R. A. Greenwald, T. Kikuchi, M. Pinnock, J. P. S. Rash, N. Sato, G. J. Sofko, J.-P. Villain, A. D. M. Walker, A decade of the Super Dual Auroral Radar Network (SuperDARN): scientific achievements, new techniques and future directions// Surveys in Geophysics, Volume 28, Issue 1, pp 33-109, 2007 
\bibitem[{{Clay and Medwin}}{1977}]{Clay_1977} Clay C.S. and Medwin H., Acoustical oceanography: principles and applications, John Willey and Sons, 1977
\bibitem[{{Givishvili et al.}}{2013}]{Givishvili_2013}Givishvili, G. V., L. N. Leshchenko, V. V. Alpatov, S. A. Grigor’eva, S. V. Zhuravlev, V. D. Kuznetsov, O. A. Kusonskii, V. B. Lapshin, and M. V. Rybakov (2013), Ionospheric Effects Induced by the Chelyabinsk Meteor// Solar System Research, 47(4), 280-287. 
\bibitem[{{Gokhberg et al.}}{2013}]{Gokhberg_2013}Gokhberg, M. B., E. V. Ol’shanskaya, G. M. Steblov, and S. L. Shalimov (2013), The Chelyabinsk Meteorite: Ionospheric Response Based on GPS Measurements// Doklady Earth Sciences, 452(1), 948-952. 
\bibitem[{{Kasahara}}{1981}]{Kasahara_1981}Kasahara K. Earthquake mechanics. Cambridge University Press, 1981. 284 p.
\bibitem[{{Klobuchar}}{1986}]{Klobuchar_1986}Klobuchar, J. A. (1986) Ionospheric time-delay algorithm for single-frequency GPS users// IEEE Transactions on Aerospace and Electronics System, 23(3), 325-331. 
\bibitem[{{Kutelev} and {Berngardt}}{2013}]{Kutelev_2014} Kutelev K.A. and Berngardt O.I., Modeling of ground scatter signal of SuperDARN radar in the presense of travelling midscale ionospheric disturbance during "Chelyabinsk" meteorite fall//Solar-terrestrial physics, V.24(137), pp.15-26, 2013 (in russian)
\bibitem[{{Maruyama} and {Shinagawa}}{2014}]{Maruyama_2014}Maruyama T. and H.Shinagawa, Infrasonic sounds excited by seismic waves of the 2011 Tohoku-oki earthquake as visualized in ionograms//Journal of Geophysical Research: Space Physics Volume 119, Issue 5, pages 4094\textendash{}4108,May 2014
\bibitem[{{Matsumura et al.}}{2011}]{Matsumura_2011}Matsumura M , Saito A., Iyemori T., Shinagawa H., Tsugawa T., Otsuka Y., Nishioka M., and Chen C.H. Numerical simulations of atmospheric waves excited by the 2011 off the Pacific coast of Tohoku Earthquake// Earth Planets Space, 63, 885–889, 2011.
\bibitem[{{Popova et al.}}{2013}]{Popova_2013}Popova O.P., P.Jenniskens, V.Emel'yanenko, A.Kartashova, E.Biryukov, S.Khaibrakhmanov, V.Shuvalov, Y.Rybnov, A.Dudorov, V.I. Grokhovsky, D.D. Badyukov, Qing-Zhu Yin, P.S. Gural, J.Albers, M.Granvik, L.G. Evers, J.Kuiper, V.Kharlamov, A.Solovyov, Y.S. Rusakov, S.Korotkiy, I.Serdyuk, A.V. Korochantsev, M.Yu. Larionov, D.Glazachev, A.E.Mayer, G.Gisler, S.V. Gladkovsky, J.Wimpenny, M.E. Sanborn, A.Yamakawa, K.L. Verosub, D.J. Rowland,S. Roeske, N.W. Botto, J.M. Friedrich, M.E. Zolensky, L.Le, D.Ross, K.Ziegler, T.Nakamura, I.Ahn, J.I.Lee, Q.Zhou, X.-H. Li, Q.-L. Li, Y. Liu, G.-Q. Tang, T.Hiroi, D.Sears, I.A. Weinstein, A.S.Vokhmintsev, A.V.Ishchenko, P.Schmitt-Kopplin, N.Hertkorn, K.Nagao, M.K.Haba, M.Komatsu, T.Mikouchi, Chelyabinsk Airburst, Damage Assessment, Meteorite Recovery, and Characterization//Science, Vol. 342 no. 6162 pp. 1069-1073, 2013. DOI: 10.1126/science.1242642
\bibitem[{{Ruzhin et al.}}{2014}]{Ruzhin_2014}Ruzhin, Yu. Ya., V. D. Kuznetsov, and V. M. Smirnov (2014) Ionospheric Response to the Entry and Explosion of the South Ural Superbolide// Geomagnetism and Aeronomy, 54(5), 601-612. 
\bibitem[{{Tauzin et al.}}{2013}]{Tauzin_2013}Tauzin, B., E. Debayle, C. Quantin, and N. Coltice (2013), Seismoacoustic coupling induced by the breakup oh the 15 February 2013 Chelybinsk meteor// Geophys. Res. Lett., 40(14), 3522-3526. 
\bibitem[{{Tertyshnikov et al}}{2013}]{Tertyshnikov_et_al_2013} Tertyshnikov A.V., Alpatov V.V., Gluhov Y.V., Perminova E.S., Davydenko D.V., Regional ionosphere disturbances and ground-based navigation positioning receiver errors with the explosion of Chelyabinsk (Chebarkul) meteoroid 15.02.2013.//Heliogeophysical Research, 2013, http://vestnik.geospace.ru/index.php?id=162
\bibitem[{{Tsugawa et al.}}{2011}]{Tsugawa_2011} Tsugawa, T., A. Saito, Y. Otsuka, M. Nishioka, T. Maruyama, H. Kato, T. Nagatsuma, and K. T. Murata, Ionospheric disturbances detected by GPS total electron content observation after the 2011 off the Pacific coast of Tohoku Earthquake// Earth Planets Space, 63, 875–879, 2011.
\bibitem[{{Ud\'{i}as et al}}{2014}]{Udias_2014}Ud\'{i}as A., Madariaga R., Buforn E. Source Mechanisms of Earthquakes: Theory and Practice // New York: Cambridge University Press, 2014. 302 p.
\bibitem[{{Yang et al.}}{2014}]{Yang_2014}Yang, Y.-M., A. Komjathy, R. B. Langley, P. Vergados, M. D. Butala, and A. J. Mannucci (2014), The 2013 Chelyabinsk meteor ionospheric impact studied using GPS measurements// Radio Science, 49(5), 341-350. 
\bibitem[{{Yuen}}{1969}]{Yuen_1969}  Yuen, P. C., P. F. Weaver, R. K. Suzuki, and A. S. Furumoto (1969), Continuous, traveling coupling between seismic waves and the ionosphere evident in May 1968 Japan earthquake data// J. Geophys. Res., 74(9), 2256–2264, doi:10.1029/JA074i009p02256.
\bibitem[{{Zotkin and Tsikulin}}{1966}]{Zotkin_1966} Zotkin, I. T. and M. A. Tsikulin, Simulation of the explosion of the Tungus meteorite// Sov. Phys. Dokl. 11, 183-186, 1966.
\end{thebibliography}
\end{document}